\documentclass[%
 aip,
 amsmath,amssymb,
 reprint,%
]{revtex4-1}

\usepackage{graphicx}
\usepackage{epstopdf}
\usepackage{amsmath}
\usepackage{multirow}
\usepackage{times}
\usepackage{amssymb}
\usepackage{dcolumn}%
\usepackage{bm}%
\usepackage{upgreek}
\usepackage{wasysym}
\usepackage{verbatim}
\usepackage{blindtext}
\usepackage[driverfallback=dvipdfm]{hyperref}
\usepackage[utf8]{inputenc}
\usepackage[T1]{fontenc}
\usepackage{mathptmx}
\usepackage[normalem]{ulem}

\hypersetup
{
	colorlinks=true,
	linkcolor=blue,
	citecolor=blue
}

\begin{document}

\preprint{AIP/123-QED}
\pagestyle{empty}

\title{Star topology increases ballistic resistance in thin polymer films}

\author{Andrea\@ Giuntoli}
\email{andrea.giuntoli@northwestern.edu}
\affiliation{Civil and Environmental Engineering Department, Northwestern University, Evanston, IL 60201, USA}

\author{Nitin K. \@ Hansoge}
\affiliation{Mechanical Engineering Department, Northwestern University, Evanston, IL 60208, USA}

\author{Sinan\@ Keten}
\email{s-keten@northwestern.edu}
\affiliation{Civil and Environmental Engineering Department, Northwestern University, Evanston, IL 60201, USA}
\affiliation{Mechanical Engineering Department, Northwestern University, Evanston, IL 60208, USA}


\begin{abstract}
\noindent
Polymeric films with greater impact and ballistic resistance are highly desired for numerous applications, but molecular configurations that best address this need remain subject to debate. We study the resistance to ballistic impact of thin polymer films using coarse-grained molecular dynamics simulations, investigating melts of linear polymer chains and star polymers with varying number ($2\leq f \leq 16$) and degree of polymerization ($10\leq M\leq 50$) of the arms. We show that increasing the number of arms $f$ or the length of the arms $M$ both result in greater specific penetration energy within the parameter ranges studied. Greater interpenetration of chains in stars with larger $f$ allows energy to be dissipated predominantly through rearrangement of the stars internally, rather than chain sliding. During film deformation, stars with large $f$ show higher energy absorption rates soon after contact with the projectile, whereas stars with larger $M$ have a delayed response where dissipation arises primarily from chain sliding, which results in significant back face deformation. Our results suggest that stars may be advantageous for tuning energy dissipation mechanisms of ultra-thin films. These findings set the stage for a topology-based strategy for the design of impact-resistant polymer films. 
\bigskip

\noindent\textit{\textbf{version accepted on Extreme Mechanics Letters}}
\end{abstract}

\maketitle

\section{Introduction}

Thin polymer films have diverse applications, but their design is challenging because of the large number of tunable parameters such as monomer chemistry, chain topology and interfaces\cite{TanakaSpecialInterfaces, SimmonsProgressFilm}. 
The interplay of these effects makes thin films very interesting to study from a fundamental point of view, where a consensus is missing even on basic properties such as the definition of the glass transition temperature, $T_g$ \cite{WengangBeyondTg}. Thin films are heavily utilized in applications such as optics, flexible electronics, coatings or packaging\cite{SorrentinoNanocoatings,MalikFilmElectronics,ClarkFilmOptics}, and advances in the mechanical design of these materials require a deep knowledge of how their mechanical properties emerge from their nanostructure. In particular, there is a growing interest in impact-resistant films and barriers for applications in airplanes and spacecrafts subjected to high-speed collisions with solid particles \cite{CarterTurbines,GrossmanSpacecrafts}, in stretchable and wearable electronics \cite{GaliotisDisplays} or for soft body armor \cite{bulletProofReview}. 

A typical method to test the impact resistance of a film is to subject the material to extreme deformation rates \cite{BackmanTarget,WenPenetration}. Small scales down to the nanoscale are now within the realm of experimental investigation thanks to laser-induced projectile impact tests (LIPIT) \cite{LeeBulletLayered,VeyssetSupersonicImpact}. These experiments quantify the specific penetration energy $E_p$ of the film from the change in the kinetic energy of the bullet, and find configurations that maximize the energy dissipation of the material while keeping its thickness, mass and production cost at a minimum. Despite recent progress, though, experiments have a hard time capturing the relaxation and dissipation mechanisms of polymer films at the molecular level and at such extreme deformation rates. Molecular dynamics (MD) simulations can support the experiments, having the potential to study them in detail in a similar setup. MD simulations can investigate the system at the femtosecond to nanosecond timescales and at high deformation rates ($10^6$ to $10^9$ $s^{-1}$), identifying the key mechanisms involved in the response to the projectile impact and guiding the experimental designs for finding optimized materials\cite{ZhaoxuBallistic,MarchiBuligandBallistic}. Star polymers, for example, show a more complex behavior in thin films when compared to linear chains\cite{GlynosVitr}, and MD simulations are able to explore the molecular mechanisms involved in the relaxation of star polymers down to the monomer level \cite{FanStars}. We expect the unique mechanical properties of star polymer melts due to their topology to be relevant for damage mitigation.

Here we investigate how a change in the topology of the polymer from linear chains to stars can lead to improved impact resistance. We perform MD simulations of coarse-grained star polymers with varying number of arms $f$ and length (degree of polymerization) of the arm $M$ and study how the change of molecular architecture influences the impact resistance under projectile impact tests. Recent literature focused on thin star polymer films, showing how fundamental properties such as $T_g$, surface absorption and free surfaces relaxation \cite{GlynosSurfaceRelaxations,WangSurfaceRelaxations,GlynosVitr,ChremosIntraStars} can be tuned by varying the molecular topology. Despite the abundant literature on the equilibrium behavior of star polymer films, in-depth studies of their mechanical properties under deformation is still missing, especially at extreme deformation rates.

\begin{figure*}
\centering
\includegraphics[width=\linewidth]{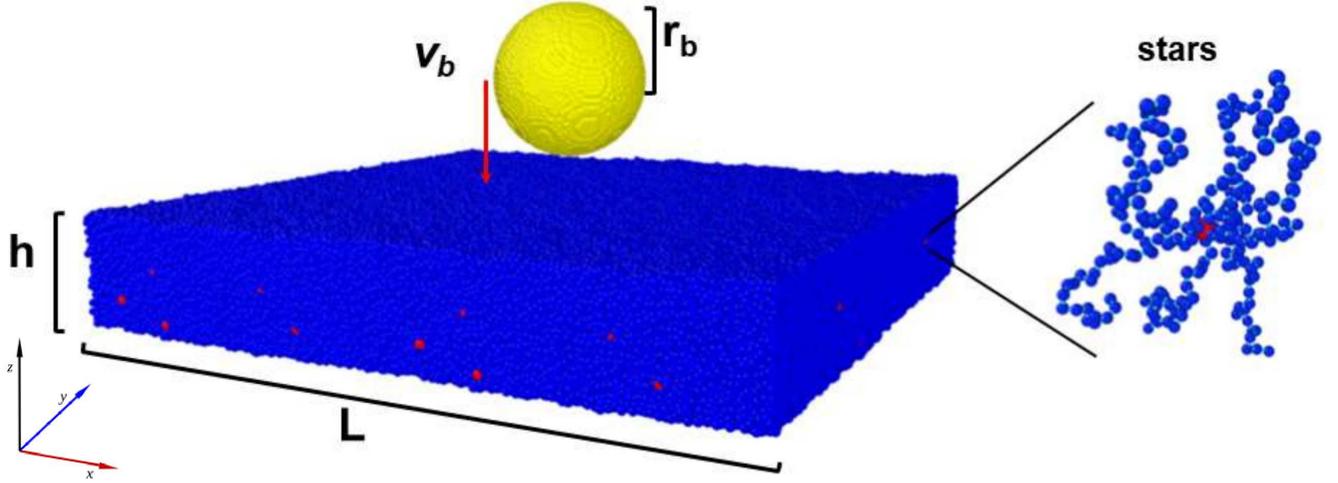}
\caption{Simulation setup of the ballistic test. The thin polymer film with thickness of $h=20$ in LJ units (corresponding to approximately $20 nm$ for polystyrene), is free standing, with no supporting interfaces. It is a melt of star polymers with varying number of arms $f$ and length of the arms $M$. A star polymer with $f=8$ and $M=30$ is shown in the figure as an example. Red particles are the star cores, while blue particles are the beads belonging to the arms. The bullet (yellow particles) is created as a spherical rigid body over the center of the film, with a radius $r_b$ ($r_b=16$ in the picture) and initial downward velocity $v_b$.}
\label{Figure1} 
\end{figure*}

In this paper, we show that increasing the number of arms $f$ of the star polymers monotonically increases the impact resistance of nanoscale thin films, producing up to a $16\%$ increase in the specific penetration energy of the film for $f=16$ arms compared to linear chains with the same arm length. We show that the chain sliding motion between different molecules plays a major role in the absorption of the projectile energy, and that this dissipation mechanism is reduced with increasing $f$ in favor of increased stretching of the bonds and intramolecular rearrangement. 
By varying the length of the arms, $M$, we also show that longer chains have a higher impact resistance, in line with recent experimental findings on polycarbonate linear chains\cite{EdwinBulletEntanglement}, but with a different mechanism. Longer chains do not show increased performance at the point of impact, but they absorb more energy over time due to the prolonged pull-out of the polymer chains as the bullet penetrates through the membrane and beyond. For ultra-thin films, this delay might not be desirable, and star polymers seem to offer advantages for the design of impact-resistant materials. These preliminary observations will pave the way for a deeper investigation of the role of molecular structure and architecture on the impact resistance of thin films.

\section{Simulation Methods}
\label{SimMet}

\subsection{Film preparation}

We prepare films of star polymers with a central bead as the nucleus and varying number of arms $f$ attached to it, where $f=2$ corresponds to linear polymer chains. Each arm has $M$ beads, so that every molecule has  $M_w=fM+1$ beads in total. For this study, we vary $f=2,4,6,8,12,16$ and $M=10,20,30,40,50$. In each system, the total number of beads is $N=130000$ or the closest number possible, when the ratio $N/M_w$ is not integer. Non-bonded monomers interact with a truncated extended Lennard-Jones (LJ) potential:

\begin{equation}
\label{eq1}
U^{LJ}(r)=\varepsilon\left [ \left (\frac{\sigma^*}{r-\Delta}\right)^{12 } - 2\left (\frac{\sigma^*}{r-\Delta}\right)^6 \right]+U_{cut} .
\end{equation}

$\Delta=0$ for the monomers in the arms, corresponding to the usual LJ potential, where $\sigma^*=2^{1/6}\sigma$ is the position of the minimum. $U_{cut}$ is chosen to ensure $U^{LJ}(r)=0$ at the cutoff distance $r \geq r_c=2.5\,\sigma$. The nucleus-nucleus interaction has $\Delta=-0.5 \sigma$ and the interaction between a nucleus and a non-bonded monomer in an arm has $\Delta=-0.25 \sigma$. This choice has the effect of having nuclei of reduced ``effective" diameter $\sigma_N=0.5\sigma$, a choice originally made to avoid self-aggregation of the cores \cite{ChremosIntraStars}.  Bonded monomers interact with a harmonic potential $U^b(r)=k(r-r_0)^2$ with $k=2500 \, \varepsilon  / \sigma^2$ and $r_0=0.9\,\sigma$. A short and stiff bond is used to avoid crystallization of the model \cite{Hanakata2015Film}, as shown before to happen for longer bonds \cite{GiuntoliStarCrystal, GiuntoliCristallo}.  $r_0=0.75 \sigma$ for the nucleus-arm bond. Harmonic bonds have been used extensively for this polymer model, even under extreme deformation rates\cite{GiuntoliSciAdv}. Bond-breaking effects are not considered here, despite their potential relevance for the localized ballistic response. Given the short chain lengths employed in this study, well below the experimental entaglement regime, we anticipate that incorporating bond-breaking should not significantly change the qualitative trends reported here. From here on, all quantities are expressed in terms of reduced Lennard-Jones units, i.e., $\varepsilon = 1$ and $\sigma=1$,  with unit monomer mass and Boltzmann constant, and unit of time $\tau=\sqrt{\epsilon/m\sigma^2}$. The reduced units can be mapped onto physical units relevant to generic non-equilibrium fluids by converting the MD time, length and energy units. For polystyrene, for example, they roughly correspond to about $9$ ps, $1$ nm and $4.1$ kJ/mol, respectively \cite{Kroger04}.

We start from a melt of star polymers in a box of lateral dimension $L_x=L_y=80$ and thickness $h=20$, with periodic boundary conditions on the lateral dimensions and fixed walls on the top and bottom of the film. The polymer beads interact with the walls with a $9-3$ Lennard-Jones potential along the $z-$axis. To reach this initial configuration, star molecules are created with a random walk algorithm and placed in a box with large initial volume. The walls are placed on top and bottom of the system, and the simulation box is squeezed to the target volume. We equilibrate the system in NPT ensemble at high temperature $T=0.75$ and pressure $P_x=P_y=10.0$ while keeping the top and bottom walls fixed, ensuring a rapid equilibration of the system. Then we release the pressure to $P_x=P_y=0$ and lower the temperature to $T=0.1$ and let the system relax. For each of these steps (high-$T$ equilibration, quench and final relaxation) we run the simulation for a time $t_{eq}=2500$ (for a total simulation time of $t=7500$). This time is chosen to allow full segmental relaxation of the polymer in the high T phase, see our SM and Figure S1 for more details. At $T=0.1$ the system is deep into the glassy state, well below the glass transition temperature $T_g$ which is around $0.3$-$0.4$ for this model in the bulk state \cite{FanStars}. At the end of the simulation the walls are removed and a short equilibration is done allowing the free surfaces to relax. Four replicas for each system are simulated to reduce statistical fluctuations. Molecular Dynamics (MD) simulations were carried out with the LAMMPS (Large-scale Atomic/Molecular Massively Parallel Simulator) code \cite{PlimptonLAMMPS}.

\subsection{Ballistic test setup}
The ballistic tests are performed by creating a spherical bullet of radius $r_b$ on top of the polymer film with an initial velocity $v_b$ along the $z-$axis. The bullet consists of beads arranged in a diamond lattice and treated as a single rigid body. The interaction between the bullet beads and the polymer beads is the same LJ non-bonded interaction described above, with $\Delta=0$. Before the start of the simulation, the polymer film is replicated once both in the $x$ and $y$ directions, resulting in a film surface four times as large, and the bullet is placed over the center of the film. The borders of the film are frozen in place to prevent the film to simply drift downward with the bullet, and only the atoms around the point of impact are allowed to move. The mobile atoms are defined by a spherical region centered in the middle of the film and with radius $r_M=70$, approximately half the final size of the simulation box, which is similar for all systems within a few percents. We study the range of $r_b=16$-$24$ and $v_b=2$-$5$ ($\sim220$-$550$ m/s for polystyrene). We verified that $r_M$ is large enough to avoid size effects due to frozen molecules being involved in the ballistic response. In particular, we see that size effects start vanishing when $r_M$ is at least twice the bullet radius, approximately. A typical setup is shown in Figure \ref{Figure1}. We measure the specific penetration energy $E_p=(K_i-K_f)/(\pi r^2_b h)$, where $K_i$ and $K_f$ are the kinetic energies of the bullet before an after the penetration, respectively, and $h$ is the film thickness. The ballistic test is performed in NVE ensemble to conserve the total energy of the system.

\section{Results and Discussion}
\label{Results}

\begin{figure}
\centering
\includegraphics[width=\linewidth]{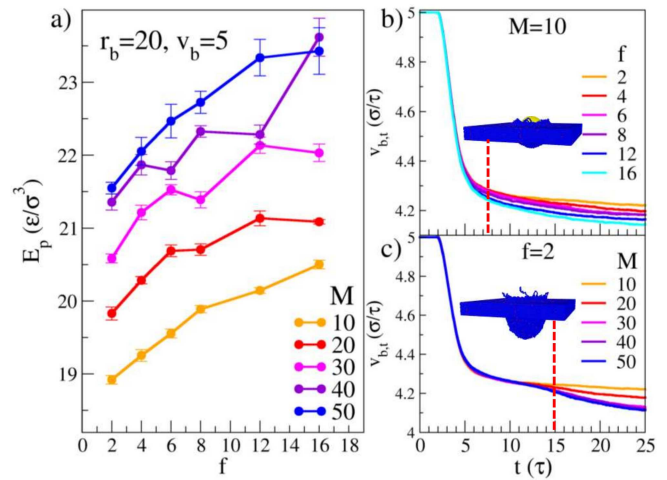}
\caption{Superior impact-resistance of star polymers and longer polymer chains. a) Specific penetration energy $E_p$ of the films with varying number of arms, $f$ and length of the arms, $M$, at fixed bullet radius and velocity $r_b=20$ and $v_b=5$. For each $M$, increasing $f$ leads to an increase of absorbed energy, proving a better performance of star polymers with respect to linear chains. Increasing $M$ also leads to increased $E_p$ across all systems, as has recently been experimentally observed for linear chains\cite{EdwinBulletEntanglement}. The observed fluctuations and outliers (see for example the data point $f=16$, $M=40$) can be explained by increased molecular interpenetration during the quench, see Figure S3. Velocity loss of the bullet in time is higher for stars with higher $f$ (b) during the penetration, while for longer linear chains (c) the increased velocity loss happens at later times, after the membrane penetration.}
\label{Figure2} 
\end{figure}

Here we present the results of the ballistic tests.  For each test, we shoot the bullet from above the center of the film, far enough that there is no interaction between the film and the bullet. The simulation runs until the bullet completely penetrates the film and moves far away from it (up to five times the film thickness). We measure the specific penetration energy absorbed by the film $E_p$ for all our system.

Figure \ref{Figure2}a shows the effect of increasing the number of arms $f$ in the star polymers for different arm lengths $M$. Irrespective of $M$, the increase of the number of arms increases the energy absorbed by the film. This observation is consistent with the assumption that the main energy dissipation mechanism comes from the activation energy required for the chains to slide over one another. The introduction of star cores and the length increase of the chains leads to increased interpenetration of different molecules. The sliding of chains thus becomes energetically more expensive, and molecular reorganization and bond stretching become favored, as we show in the following. Different curves in Figure \ref{Figure2}a also show that an increased chain length $M$ provides better impact resistance, a result recently observed experimentally for linear chains \cite{EdwinBulletEntanglement} and that we retrieve for varying degree of number of arms $f$. It is important to note that the times at which the effect of $f$ or of $M$ comes into play are different. Figure \ref{Figure2}b and \ref{Figure2}c show the velocity loss of the bullet in time for varying $f$ or $M$, respectively. Films with linear longer chains have increased $E_p$ at later times ($t\sim15$ in Figure \ref{Figure2}c), pulling on the projectile which already went beyond the thin membrane. Star polymers with increasing $f$ thus show better impact-resistance at earlier stages, which is particularly relevant for ultra-thin films applications. Similar trends are observed for the whole range of bullet radius and velocity, and we report our full data in the Supplementary Material (see Figure S2).
We notice some fluctuations and outliers in our data, (see the data point for $f=16$, $M=40$, for example) which are due to systems that reached a state with increased molecular interpenetration. Interestingly, we could observe this feature from the energy of the pairwise interactions between different molecules, but from no other marker (pressure, density, global potential energy, etc...). A full report of our observations can be found in Figure S3.

\begin{figure}
\centering
\includegraphics[width=\linewidth]{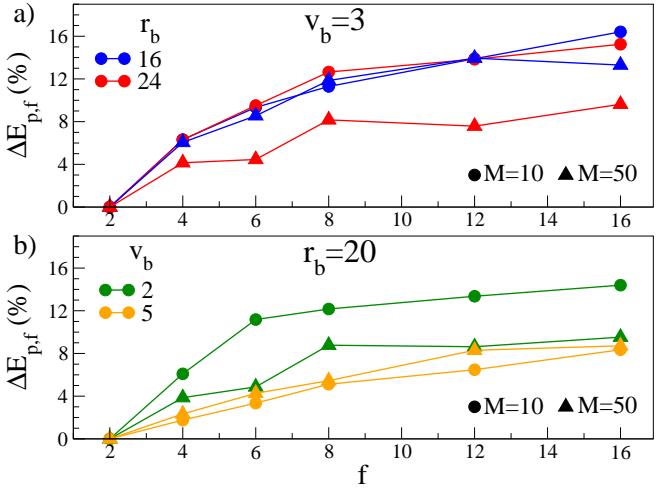}
\caption{Percentage gain $\Delta E_{p,f}$ in the specific penetration energy $E_p$ when moving from linear chains to $f=16$ star polymers for the whole range of bullet parameters investigated varying bullet radius, panel a), or bullet velocity, panel b). The lowest and highest lengths of the arms are shown, $M=10$ (circles) and $M=50$ (triangles). A consistent increase between $\sim7\%$ and $\sim16\%$ is observed for $f=16$. Maximum gain is observed for the lowest arm length $M=10$, where the difference in polymer architecture is more relevant. This is more apparent for larger bullet radius $r_b$ and lower bullet velocity $v_b$. Full data reported in FigureS4.}
\label{Figure3} 
\end{figure}

We are interested in quantifying how much is gained by creating star polymers with respect to linear chains. We then introduce the quantity $\Delta E_{p,f}=(E_{p,f}-E_{p,2})/E_{p,2}$, where $E_{p,2}$ and  $E_{p,f}$ are the specific penetration energies of linear chains and star polymers with $f$ number of arms, respectively, at fixed $M$. $\Delta E_{p,f}$ is then the percentage gain in absorbed energy when moving from linear chains to star polymers with varying $f$. We expect $\Delta E_{p,f}$ to be higher for lower $M$, where the presence of a core particle is more relevant, and for bullet parameters that are closer to the length and time scales of the polymer system, presumably the gyration radius $R_g$ and segmental relaxation time $\tau_\alpha$,  where the different polymer architecture can play a role in the penetration and dissipation process.

Figure \ref{Figure3} reports $\Delta E_{p,f}$ as a function of $f$ for fixed bullet velocity $v_b$ and varying bullet radius $r_b$ (panel a) or for fixed $r_b$ and varying $v_b$ (panel b). We show the extremes of the parameters $M$, $r_b$ and $v_b$ investigated. At a first order we observe that an improvement of $8-16\%$ is present for all our systems when moving to $f=16$. A larger gain is observed for $M=10$, as expected. The difference between $M=10$ and $M=50$ is more evident for larger bullets, where more polymer molecules are involved in the impact, and for lower velocities, presumably at time scales where the polymer segmental motion and the stars interdigitation are more relevant to the relaxation and dissipation mechanisms of the thin film. We believe that a more quantitative analysis can be derived taking into account the different length and time scales involved, such as bullet velocity and radius compared to the film thickness and the polymer size and relaxation time, but at the present stage we do not observe obvious trends and more analyses and in-depth studies are needed.

\begin{figure}
\centering
\includegraphics[width=\linewidth]{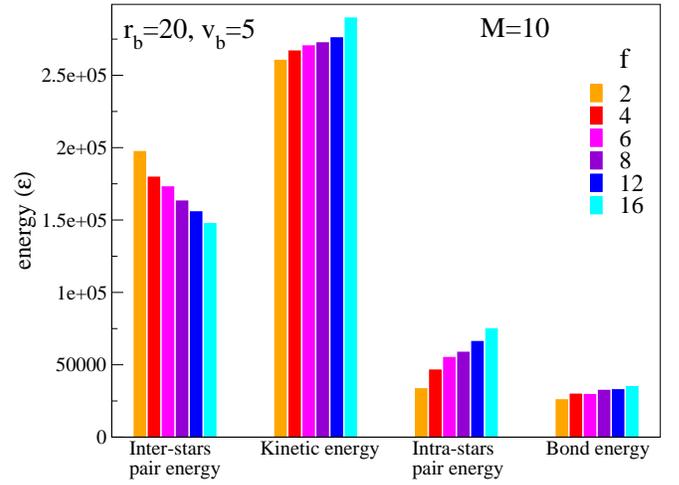}
\caption{Energy changes of the polymer film during impact, for $r_b=20$ and $v_b=5$ varying $f$, with $M=10$. The energy lost by the bullet and absorbed by the film is decomposed into inter-stars pairwise interactions, kinetic energy, intra-stars pairwise interactions and bonds interactions, reported with respect to their equilibrium values. With increasing $f$, less energy goes into inter-stars interactions and more energy goes into kinetic and intra-stars contributions, leading to an overall better performance of the film. The same trends are observed for all our systems, see Figure S5.}
\label{Figure4} 
\end{figure}

In Figure \ref{Figure4} we provide more insight into the dissipation mechanisms of our stars films. We report the changes in energy of the polymer film during the impact, decomposed in several parts, for the bullet with $r_b=20$ and $v_b=5$ as in Figure \ref{Figure2}, for $M=10$ and varying $f$. The change in pairwise energy between \textit{different} stars, panel (a), is lower with increasing $f$, meaning that larger star molecules, being more intertwined, are sliding less one on the other during the impact. To compensate that, at higher $f$ more energy goes into kinetic energy of the film, intra-stars pairwise energy, and, to a lesser degree, into stretched bonds. The contribution of these different pieces leads to an overall increase in the specific penetration energy of the film, as shown in Figure \ref{Figure2}. 
Similar trends are observed for all our systems both with increasing $f$ and $M$, and we report more complete data in Figure S5. We infer that with increasing $f$ (and thus decreasing number of molecules, being the total number of beads fixed) the larger amount of \textit{internal} interactions (bonded and non-bonded) plays a larger role in the dissipation of the bullet energy. Internal molecular rearrangement is favored with respect to chain sliding from different molecules. This leads to whole star molecules being stretched and dragged (thus the increased kinetic energy), and ultimately to an overall increase in the material performance. We also note that the change in bond energy shown in Figure \ref{Figure4} is $48\%$ higher for $f=16$ than for $f=2$, while the number of bonds in the mobile region is only $3.7\%$ higher, so the increase in bond energy change cannot be attributed simply to the difference in the total number of bonds.
Different advantages at different time and length scales are provided by either increasing the chain length or the complexity of the polymer topology, and we envision that our preliminary observations will pave the way for a larger investigation into the design of impact-resistant polymer films taking advantage of the polymer topology.

\section{Conclusions}
There is a growing interest in impact-resistant materials, and thin films in particular, but a theoretical knowledge of the mechanisms involved in energy dissipation is still lacking and there is vast room for improvement in the design space of new materials using computations. In the present paper, we show that an increase in complexity of the molecular architecture can increase the impact resistance of thin polymer films with minimal effort. We simulate thin polymer films composed of melts of star polymers with varying number of arms $f$ and length of the arms $M$. We show that the energy absorbed by the film during a projectile-impact test at extreme deformation rates increases with increasing number of arms. We report an increased performance of up to $16\%$ between star polymers with large $f$ and linear polymer chains, paving the way for a topology-controlled protocol to increase the impact resistance of thin polymer films.
In general, we observe that for both higher $f$ and $M$, increasing the molecular weight and size of the molecules, less energy goes into intermolecular exchanges during the projectile penetration. Conversely, more energy is stored into intramolecular rearrangements and material kinetic energy, which we associate with a higher polymer mass being extruded from the film during the impact. While longer linear chains have higher penetration energy because they keep pulling on the bullet after the penetration of the film, star polymers have immediate higher stopping power, an advantage for ultra-thin films applications.  More in-depth studies of our finding will allow to tune and optimize these mechanisms for varying impact conditions (like different radius or velocity of the projectile). These investigations will deepen our understanding of the role of polymer topology under extreme deformations and will pave the way for applications such as improved body armors and impact-resistant coatings. 

\section*{Acknowledgements}
The authors acknowledge funding from the Army Research Office (award  W911NF1710430). We have no conflict of interest to declare.

\bibliography{biblio} 

\end{document}